\documentclass[journal]{IEEEtran}
\usepackage{amsfonts}
\usepackage[fleqn]{amsmath}
\usepackage{algorithm}
\usepackage{algorithmic}
\usepackage{array}
\usepackage[caption=false,font=normalsize,labelfont=sf,textfont=sf]{subfig}
\usepackage{textcomp}
\usepackage{stfloats}
\usepackage{multirow}
\usepackage{multicol}
\usepackage{url}
\usepackage{verbatim}
\usepackage{graphicx}
\usepackage{rotating}
\usepackage{tabularray}
\usepackage{pbox}
\usepackage{makecell}
\usepackage{footnote}
\usepackage{nicematrix}
\usepackage{booktabs}     
\usepackage{graphicx}     
\usepackage{booktabs}
\usepackage{siunitx}
\usepackage{multirow}
\usepackage{array}
\usepackage{xcolor}
\usepackage{amsmath, amssymb, mathtools}
\usepackage{multirow,arydshln,booktabs} 
\def\BibTeX{{\rm B\kern-.05em{\sc i\kern-.025em b}\kern-.08em
    T\kern-.1667em\lower.7ex\hbox{E}\kern-.125emX}}
\usepackage{balance}

\usepackage[style=ieee, maxnames=2, minnames=1, url=false, doi=false, isbn=false]{biblatex}
\AtEveryBibitem{\clearfield{langid}}

\AtEveryBibitem{\clearlist{language}}

\AtEveryBibitem{
  \clearfield{note}
  \clearfield{addendum}
  \clearfield{howpublished}
  \clearfield{language}}

\allowdisplaybreaks

\begin{document}
\bstctlcite{BSTcontrol}
\title{Missing Money and Market-Based Adequacy in Deeply Decarbonized Power Systems with Long-Duration Energy Storage}
\author{Adam Suski,~\IEEEmembership{Graduate Student Member, IEEE}, Elina Spyrou,~\IEEEmembership{Member, IEEE}, Richard Green
\thanks{This work was funded by the Taylor Donation from the Grantham Institute and Energy Futures Lab, Imperial College London, the Leverhulme International Professorship with grant reference LIP-2020-002, and the Engineering and Physical Sciences Research Council under the grant EP/Y025946/1 (Electric Power Innovation for a Carbon-free Society (EPICS)).}
\thanks{Adam Suski and Elina Spyrou are with the Department of Electrical and Electronic Engineering, Imperial College London. (e-mail: a.suski23@imperial.ac.uk). Richard Green is with the Department of Economics \& Public Policy, Imperial College Business School.}}

\maketitle

\begin{abstract}
The ability of deeply decarbonised power systems to ensure adequacy may increasingly depend on long-duration energy storage (LDES). A central challenge is whether capacity markets (CMs), originally designed around thermal generation, can provide efficient investment signals when storage becomes a central participant. While recent studies have advanced methods for accrediting variable renewables and short-duration storage, the effectiveness of these methods in CMs with substantial LDES penetration remains largely unexplored. To address this gap, we extend a two-stage stochastic equilibrium investment model by endogenising continuous, duration-based capacity accreditation for storage and apply it to a Great Britain-based case using 40 years of weather-driven demand and renewable profiles under varying emission limits. Results show that well-calibrated CMs can sustain near-efficient investment and mitigate revenue volatility, but their effectiveness diminishes in deeply decarbonized systems, underscoring both their potential and the regulatory challenges of supporting large-scale LDES.
\end{abstract}

\begin{IEEEkeywords}
Long-Duration Energy Storage, Missing Money Problem, Capacity Market, Capacity Accreditation
\end{IEEEkeywords}

\section*{Nomenclature}
\addcontentsline{toc}{section}{Nomenclature}

\subsection*{Sets}
\begin{IEEEdescription}[\IEEEusemathlabelsep\IEEEsetlabelwidth{$\mathcal{T}, \mathcal{G}, \mathcal{S}$}]
\item[$\Omega$] Set of scenarios.
\item[$\mathcal{T}$] Set of time periods.
\item[$\mathcal{G}$] Set of generators.
\item[$\mathcal{S}$] Set of storage units.
\item[$\mathcal{R}$] Set of resources $\mathcal{G}\cup \mathcal{S}$.
\item[$\mathcal{Z}$] Set of durations.

\end{IEEEdescription}
\vspace{-3mm}

\subsection*{Variables}
\begin{IEEEdescription}[\IEEEusemathlabelsep\IEEEsetlabelwidth{$q_{\omega t s}^{ch}, q_{\omega t s}^{dis}$}]
\item[$c_{g}, c_{s}^{P}, c_{s}^{E}$] Installed capacities: generator, storage power, storage energy.
\item[$c^{CM}_g, c^{CM}_s$] Capacity market bids: generator, storage.
\item[$q_{\omega t g}$] Generator power generation.
\item[$q_{\omega t s}^{\text{ch}}, q_{\omega t s}^{\text{dis}}$] Storage charging and discharging power.
\item[$d_{\omega t}^{\text{fix}}, d_{\omega t}^{\text{flex}}$] Demand met: price inelastic, flexible.
\item[$e_{\omega t s}, e_{s}^{\text{init}}$] Storage energy: at time $t$ and initial.
\item[$d^E_{\omega, t}, d^{C}$] Consumption of energy and capacity .
\item[$\lambda^E_{\omega,t}, \lambda^{CM},$] Energy, capacity and emission price
\item[$\lambda^{CT}$] 
\end{IEEEdescription}
\vspace{-3mm}

\subsection*{Input Parameters}
\begin{IEEEdescription}[\IEEEusemathlabelsep\IEEEsetlabelwidth{$\widetilde{I}_{s}^{E}, I_{s}^{E.F}$}]
\item[$w_{\omega t}$] Duration of time period [hours].
\item[$\delta_{\omega}$] Probability weight of scenario.
\item[$B$] Benefit factor for meeting demand.
\item[$D_{\omega t}^{\text{fix}}, D_{\omega t}^{\text{flex}}$] Maximum demand: fixed, flexible.
\item[$C_{g}^{V}$] Variable cost of generator.
\item[$\widetilde{I}_{g}, I_{g}^{F}$] Generator costs: investment, fixed operational.
\item[$\widetilde{I}_{s}^{P}, I_{s}^{P.F}$] Storage power capacity costs: investment, fixed operational.
\item[$\widetilde{I}_{s}^{E}, I_{s}^{E.F}$] Storage energy capacity costs: investment, fixed operational.
\item[$\eta_{s}^{\text{ch}}, \eta_{s}^{\text{dis}}$] Storage efficiencies: charging, discharging.
\item[$A_{\omega g t}$] Availability factor of generator.
\item[$\text{EF}_g$] Emission factor of generator $g$.
\item[$\Bar{E}^{CT}$] Annual expected emission limit.
\item[$\text{VOLL},\text{PC}$] Value of lost load, price cap
\end{IEEEdescription}

\section{Introduction}
\IEEEPARstart{R}{ecent} long‐term planning studies have underscored the indispensable role of large‐scale energy storage (ES) in achieving deep decarbonisation of the power sector \cite{jafari_power_2020}. Modelling analyses consistently suggest that in futures with very \textcolor{black}{low} greenhouse gas (GHG) emission targets, the additional energy capacity (MWh) requirement is much higher relative to the additional power capacity (MW), indicating that investments in long duration energy storage (LDES) technologies\footnote{While LDES does not have a common definition, it usually refers to a class of storage technologies with discharge duration that exceeds 6 hours \cite{twitchell_defining_2023}.} will be essential \cite{cebulla_how_2018}. This need becomes particularly acute in the full or partial absence of alternative clean, firm resources, such as new nuclear power or natural‐gas plants equipped with carbon capture and storage (CCS) \cite{de_sisternes_value_2016}.

Integration of LDES faces at least two challenges. First, many LDES technologies are capital-intensive and relatively immature. To guide technology development, a recent research stream explored cost thresholds for LDES to appear in least-cost capacity mixes, concluding that significant reductions in energy-capacity costs are needed to compete with other firm capacity resources \cite{sepulveda_design_2021}. Second, the business models that allow LDES to valorize all its operational and environmental benefits for the system \cite{cole_peaking_2023} have not yet been established. In energy markets with price caps lower than the value of lost load (VOLL), a business model purely based on energy arbitrage remains uneconomic, with narrow price spreads failing to yield sufficient returns to fully support investments \cite{schmidt_monetizing_2023}.
 
Pure arbitrage is equally unattractive for short-duration energy storage (SDES), e.g., electrochemical batteries. However, in the case of SDES, stacking revenues from multiple market products such as energy and ancillary services \cite{tian_stacked_2018}, combined with generous subsidies, such as tax credits or grants \cite{dougherty_improving_2021}, has boosted batteries' profits and deployment. These two drivers have created robust SDES business models, leading to a sixfold surge of global annual installations since 2020 \cite{iea_batteries_2024}. However, these models will likely not be transferable to LDES. Short-term applications, such as frequency regulating reserves and diurnal arbitrage, will \textcolor{black}{likely} continue to be dominated by SDES and additionally face market saturation risks \cite{pollitt_competition_2021}. Furthermore, subsidies and grants can become unsustainable if a large number of clean energy assets need support.

One benefit of LDES lies in its peaking potential and its contribution to system adequacy \cite{cole_peaking_2023}. As such, in decarbonized power systems, LDES  primarily replaces dispatchable conventional peakers, limits renewable curtailment, and ensures adequacy during contingencies or prolonged periods with low‑VRE generation (so‑called `energy droughts' \cite{bracken_standardized_2024}). Consequently, as with peaking units, a substantial portion of the LDES revenues is likely to be realized during scarcity or near-scarcity events. In markets with price caps lower than the VOLL, insufficiently high price spikes during scarcity events could create a missing money problem \cite{newbery_missing_2016}, challenging the viability of investment in LDES. Furthermore, the concentration of scarcity events within particular weather years can amplify interannual profit variability, undermining investor confidence even when wholesale energy price arbitrage provides sufficient revenue over the financial lifetime of the asset \cite{mallapragada_electricity_2023}.

Capacity remuneration mechanisms (CRMs), such as capacity markets (CMs), address missing money and revenue volatility in price‑capped markets. However, CMs were originally designed with thermal generation in mind, and their effectiveness in contexts with substantial ES capacity, particularly LDES, remains understudied. Recent research has focused on accurately valuing the contribution of energy storage to system adequacy.  Zachary et al. \cite{zachary_integration_2022} demonstrated that a change in expected unserved energy (EUE) is a suitable metric for evaluating the contribution of resources to system adequacy, extending the analyses to VRE and storage (yet with strong assumptions about its dispatch). They also demonstrated that a marginal, rather than an average, approach could yield accurate investment signals\footnote{Marginal capacity credit reflects the incremental contribution of a new resource to system adequacy, while average capacity credit captures the overall contribution of all existing resources of that type.}.  Although marginal accreditation methods are more complex to apply, multiple CM operators, including NYISO \cite{mohrman_capacity_2022}, ISO-NE \cite{zhao_resource_2022}, PJM \cite{noauthor_order_2024}, and GB \cite{national_energy_system_operator_storage_2024}, have recently begun to progress in this direction.

The shift towards marginal accreditation is promising, as investment decisions are highly sensitive to the chosen methodology and underlying assumptions \cite{mertens_capacity_2021}. Yet, even when marginal credits are used for ES and VRE, there is no general proof ensuring that investments will be efficient. Wang et al. \cite{wang_crediting_2022} demonstrated through simulations that marginal accreditation in CMs can provide price signals that encourage near-efficient investment for VRE and SDES. While these results are encouraging for the effectiveness of CMs in the near future, with narrow participation of ES, they are of limited use for future systems where LDES may play a dominant role.

In general, most existing studies focus on ES with a single value (usually short) of duration or study systems where ES is a minor contributor to a system's resource adequacy \cite{wang_crediting_2022, opathella_novel_2019}. Many that include ES in CRMs often make simplified assumptions about accreditation \cite{ sanchez_jimenez_capacity_2024}. These simplifications make the results of existing studies less relevant for systems with LDES. The duration of LDES, for many emerging technologies, is a design choice independent of the power capacity. Moreover, LDES duration can range from a few to several hundred hours.  Different durations yield distinct contributions to system adequacy, complicating decisions for both investors and regulators. 

Given the lack of studies evaluating the effectiveness of CMs in providing efficient investment signals for LDES, this article fills this gap by simulating a system with LDES and assessing CM performance, contrasting their results with those of a competitive market and no price cap. We develop an equilibrium model for the electricity market that makes investment and dispatch decisions under a cap on CO\textsubscript{2} emissions. We advance current practice by developing and including in our model an accreditation mechanism that explicitly considers capacity credits as a function of ES duration. Our evaluation framework assesses how CMs mitigate the missing‑money problem in terms of both welfare and adequacy. In particular, our study addresses the following questions:

\begin{itemize}
  \item How significant is the missing money problem for LDES in price-capped electricity markets under different decarbonisation targets?
  \item To what extent can well-calibrated CMs, incorporating marginal accreditation, support efficient investment outcomes?
  \item How do inaccuracies in marginal accreditation affect market outcomes and adequacy?
  \item To what extent do CMs reduce interannual volatility in energy prices and LDES revenues under different decarbonisation targets?
\end{itemize}

We contribute to the literature in several ways. Methodologically, we extend the framework of \cite{wang_crediting_2022} by incorporating a price-elastic demand to address degeneracy and improve price interpretability under high VRE and storage penetration. \textcolor{black}{This is also the first study to endogenise continuous ES accreditation within an investment model, using piecewise linear duration functions. Unlike previous studies (e.g., \cite{mertens_capacity_2021}, \cite{askeland_equilibrium_2019}) that estimate capacity credits for a discrete set of predefined duration levels, the proposed continuous ES accreditation curves provide capacity credits for a continuous set of durations, allowing the model to choose optimal storage configurations.}

Our analysis offers new insights into market-based adequacy in low-carbon systems with high LDES levels. We confirm that the missing money problem remains significant under price caps, but declines as storage is able to capture wider spreads \textcolor{black}{by avoiding VRE curtailment}. We also show that the price cap affects revenues and operations below the cap itself, particularly in cases of low CO\textsubscript{2} emissions. Well-calibrated CMs with marginal EUE-based accreditation maintain near-efficient investment signals, but their effectiveness is lower under \textcolor{black}{lower} decarbonization targets.

These findings are relevant for CM design and resource adequacy methodologies. They also inform policymakers considering LDES support mechanisms, such as the UK’s cap-and-floor scheme \cite{department_for_energy_security__net_zero_long_2024} by highlighting the interplay between accreditation, revenue volatility, and revenue sufficiency.

The rest of this paper is structured as follows. Section II describes the methodology, Section III presents the case study, Section IV discusses the results, and Section V concludes with a synthesis of our findings and suggestions for future research.

\section{Methodology}

This section presents the equilibrium model, including its mathematical formulation, solution strategy, capacity accreditation method, CM parameterisation, and experimental design.

\subsection{Equilibrium capacity expansion model}\label{subs:Model}

Here, we introduce a two-stage stochastic equilibrium model for the investment and dispatch in the electricity market. In the first stage, investments are made, and the capacity market is cleared, and the emission price is decided through a cap-and-trade CO\textsubscript{2} scheme. In the second stage, dispatch is decided when the energy market is cleared. To capture variations in load profiles and VRE availability, the second stage simulates dispatch decisions for alternative annual scenarios represented by the set $\Omega$. We model three types of participants: generators, ES operators, and consumers, all of whom are assumed to bid truthfully in a perfectly competitive environment. Although the current setup disregards the network, it can be readily extended to include the system operator. 

\subsubsection{Generators}
Let \(\mathcal{G}\) be a set of electricity generators. Each generator has three decision variables: its installed capacity (\(c_g\)), the capacity contracted by the CM  (\(c^{\text{CM}}_g\)), and the electricity produced in each time interval (\(q_{\omega t g}\)), which are in the set \(\Gamma^g=\left\{q_{\omega t g},c_g,c^{\text{CM}}_g \geq 0 \right\}\). Acting as a price taker, each generator solves the optimization problem \eqref{eq:Gen} below.
\begin{subequations}
\label{eq:Gen}
\begin{gather}
\forall g \in \mathcal{G} \quad \max_{\Gamma^{g}} \sum_{\omega t } w_{\omega t} \delta_{\omega} \left(\lambda^E_{\omega t}-C_g^V - \lambda^{\text{CT}} \text{EF}_g\right)q_{\omega t g} \notag \\
+ c^{\text{CM}}_g\lambda^{\text{CM}} - c_{g}\left( \widetilde{I}_{g} + I_{g}^{F}\right) \hspace*{\fill} \label{eq:Gen_1} \\
q_{\omega t g} \leq c_{g} A_{\omega t g} \hspace*{\fill} \quad \forall  \omega, t; \label{eq:Gen_2} \\
c^{\text{CM}}_g \leq \text{CC}_gc_{g} \hspace*{\fill} \label{eq:Gen_3}
\end{gather}
\end{subequations}
Eq. \eqref{eq:Gen_1} is the generators' objective function, maximising the expected net revenue from participation in the three markets, reduced by the operational and annualised investment costs. Eq. \eqref{eq:Gen_2} limits hourly production to the installed capacity multiplied by the hourly capacity factor, while Eq. \eqref{eq:Gen_3} limits the CM capacity to the accredited capacity.
\subsubsection{Storage}
Let set $\mathcal{S}$ include all ES market participants. ES participants determine power and energy capacity independently. Additionally, ES decision variables include the capacity contracted by the CM, its initial state-of-charge (SOC), its charging, and discharging. All decision variables are in $\Gamma_s=\left\{q_{\omega,t,s}^{\text{dis}},q_{\omega,t,s}^{ch},c_s^P,c_s^E,e_{\omega,t,s},e^{\text{init}},c^{\text{CM}}_s \geq 0\right\}$ and their values are determined by solving the optimization problem (2).
\begin{subequations}
\label{eq:Stor}
\begin{gather}
\forall s \in \mathcal{S} \quad \max_{\Gamma^{s}} \sum_{\omega t} w_{\omega t} \delta_{\omega} \lambda^E_{\omega,t}\left(q_{\omega t s}^{\text{dis}}-q_{\omega t s}^{\text{ch}}\right) \notag \\
+ c^{\text{CM}}_s\lambda^{\text{CM}}  -c_s^P\left({\widetilde{I}}_s^P+I_s^{P.F}\right)-c_s^E\left({\widetilde{I}}_s^E+I_s^{E.F}\right)\hspace*{\fill}  \label{eq:Stor_1}\\
q_{\omega t s}^{\text{dis}} \leq c_{s}^{P} \hspace*{\fill} \quad \forall \; \omega , t  \label{eq:Stor_2}\\
q_{\omega t s}^{\text{ch}} \leq c_{s}^{P} \hspace*{\fill} \quad \forall \; \omega, t \label{eq:Stor_3}\\
e_{\omega t s} = e_{\omega t-1 s} + w_{\omega t}(q_{\omega t s}^{\text{ch}}\eta_{s}^{\text{ch}} - \frac{q_{\omega t s}^{\text{dis}}}{\eta_{s}^{\text{dis}}}) \quad \forall \; \omega ,\;t>1 \label{eq:Stor_4}\\
e_{\omega t s} = e_s^{\text{init}} + w_{\omega t}(q_{\omega t s}^{\text{ch}}\eta_{s}^{\text{ch}} - \frac{q_{\omega t s}^{\text{dis}}}{\eta_{s}^{\text{dis}}}) 
\quad \forall \; \omega,\; t=1 \label{eq:Stor_5}\\
e_{\omega t s} \geq e_s^{\text{init}} \hspace*{\fill} \; \forall \; \omega , \;  t = T \label{eq:Stor_6} \\
e_{\omega t s} \leq c_{s}^{E} \hspace*{\fill} \quad \forall \; \omega ,  t   \label{eq:Stor_7}  \\
c^{\text{CM}}_s \leq c_{s}^{P} f^{\text{CC}}(c_{s}^{E},c_{s}^{P}) \hspace*{\fill} \label{eq:Stor_8}
\end{gather}
\end{subequations}

Eq. \eqref{eq:Stor_1} is the ES's objective function, maximising its profit. Eq. \eqref{eq:Stor_2}-\eqref{eq:Stor_3} limit charging and discharging based on installed power capacity. Eq. \eqref{eq:Stor_4}-\eqref{eq:Stor_5} calculate the energy stored at the end of a time interval of duration $w_{\omega t}$ by considering the energy stored at the start of the interval and the charging, and discharging during the interval. Eq. \eqref{eq:Stor_6}-\eqref{eq:Stor_7} limit the energy stored relative to the energy stored at the start of the model's horizon and the installed energy capacity, respectively\footnote{Constraining end-of-horizon \textcolor{black}{stored energy} to exceed initial levels may be restrictive, but it provides a practical way to handle model end effects under limited foresight. It is also consistent with EU rules requiring minimum gas storage levels before winter \cite{european_comission_amending_2025}.}. 

Finally, \eqref{eq:Stor_8} limits the capacity contracted by the CM, \( c^{\text{CM}}_s \), to the qualifying capacity of ES. We determine the 
capacity credit of ES in a novel manner as a function, \( f^{\text{CC}} \), of ES duration $\zeta_s$ \(  = \eta_s^{	\text{dis}}c_s^E/c_s^P \). This new way allows us to properly consider the capacity market contributions of a range of ES durations. The function \( f^{CC} \) is piece-wise linear with $|L|$ segments, where each segment \( l \) covers a range of durations \( [\zeta_{sl-1}, \zeta_{sl}] \). As such, \( f^{\text{CC}}\) is represented as 
\begin{equation}
f^{\text{CC}}(\zeta_s) = \alpha_{ls} + \beta_{ls} \zeta_s, \quad \forall \; \zeta_s \in [\zeta_{sl-1}, \zeta_{sl}], l\in\mathcal{L}
\end{equation}  

Capacity credit curves are monotonically increasing and concave with respect to duration \cite{paul_denholm_beyond_2023}. In other words, the marginal contribution to adequacy diminishes as duration increases (i.e., the value of $\beta_{l s}$ is positive but lower for higher durations).
We estimate  the coefficients  of \( f^{\text{CC}}\), i.e., $\alpha_{l s},\beta_{l s} $, using piece-wise linear regression. 

In the maximisation problem (2), constraint \eqref{eq:Stor_8} can be included as is because \( f^{\text{CC}}\) is concave. If \( f^{\text{CC}}\) was not concave, binary variables and associated constraints would be needed to ensure that the segments of the piecewise curve are activated in a logical order. Multiplying both sides of (3) with the installed power capacity $c_s^P$,  \eqref{eq:Stor_8} becomes: 
\begin{equation}
c^{\text{CM}}_s \leq \alpha_{l s} c_s^P + \beta_{l s}\eta_s^{\text{dis}}c_s^E, \quad \forall l\in\mathcal{L}
\end{equation}

\subsubsection{Consumers}
We assume a single representative agent acting on behalf of all consumers. Its decision variables include the electricity consumption and capacity purchased in the CM, grouped in the set $\Gamma^{\text{CM}} = \left\{\left.d^E_{\omega t}\right|t\in\mathcal{T}, \omega\in\Omega, d^{\text{CM}} \geq 0 \right\}$. Consumers maximise the expected consumer surplus, which is defined as the difference between the integral of willingness to pay (WTP) functions for energy $f^{E}$ and capacity $f^{\text{CM}}$ and the procurement cost (product of price and quantity purchased).
\begin{subequations}
\begin{multline}
\label{eq:Consumer_WTP}
\max_{\Gamma^{C}} \sum_{\omega t} w_{\omega t} \delta_{\omega} \Bigg[ \int_{0}^{d^E_{\omega  t}}f^E_{\omega t}(x)dx -\lambda^E_{\omega  t}d^E_{\omega t}\Bigg]\\
 + \int_{0}^{d^{\text{CM}}}f^{\text{CM}}(x)dx - \lambda^{\text{CM}}d^{\text{CM}} \hspace*{\fill}
\end{multline}
\end{subequations}

The demand curves for electricity can take diverse shapes, ranging from perfectly price-inelastic to perfectly price-elastic curves. In this work, in line with \cite{de_jonghe_optimal_2012}, we assume that the demand curve for electricity at each time interval has two segments. \textcolor{black}{The first segment is perfectly elastic (horizontal) for the willingness to pay for electricity up to the quantity demanded of $D^{\text{fix}}_{\omega t}$ is constant at $B$.} The second segment is a price-elastic linear demand curve, with willingness to pay ranging from $0$ to $B$ for consumption between $0$ and $D^{\text{flex}}_{\omega t}$. This demand curve helps ensure price uniqueness and resolve dual degeneracy issues in systems with high shares of renewables and storage. The WTP function for energy $f^E$ is included in the model using the equations below.
\begin{subequations}
\label{eq:Consumer}
\begin{gather}
\int_{0}^{d^E_{\omega t}}f^E_{\omega t}(x)dx = B \left( d_{\omega t}^{\text{fix}} + d_{\omega t}^{\text{flex}} - \frac{ (d_{\omega t}^{\text{flex}}) ^{2}}{2D_{\omega t}^{\text{flex}}} \right) \hspace*{\fill} \label{eq:Consumer1}\\
d_{\omega t}^{\text{fix}} \leq D_{\omega t}^{\text{fix}} \quad \forall \omega t \hspace*{\fill} \label{eq:Consumer2} \\
d_{\omega t}^{\text{flex}} \leq D_{\omega t }^{\text{flex}} \quad \forall  \omega, t  \hspace*{\fill} \label{eq:Consumer3}
\end{gather}
\end{subequations}

The CM function is represented similarly and often has more segments with varying levels of elasticity. The CM function $f^\text{CM}(d^\text{CM})$ represents the WTP for capacity, and is a monotonically decreasing function. It consists of $N$ segments, with differing levels of elasticity. The first segment is perfectly price-elastic \textcolor{black}{(horizontal)}, willing to pay up to the capacity market price cap $B^C_1$ for any capacity between 0 and $D^{\text{C-fix}}$. In the last segment, the WTP $B^C_N$ decreases to zero as $d^C$ reaches the maximum capacity the CM aims to procure. The $N-2$ segments in between are flexible (lengths $D_n^{\text{C-flex}}$, starting prices $B^C_n$, with $B^C_N = 0$) and they are parametrized based on the net cost of new entry (CONE). Given this structure, the function can be represented as:
\begin{multline}
\int_{0}^{d^{\text{CM}}=d^{\text{fix}} + \sum_{n=1}^{N-1}d_n^{\text{flex}}} f^{\text{CM}}(x) \, dx = B^C_1 d^{\text{C-fix}} + \\
\sum_{n=1}^{N-1} B^C_n d_n^{\text{C-flex}} - 
 \sum_{n=1}^{N-1} \frac{(d_n^{\text{C-flex}})^2 (B^C_n - B^C_{n+1})}{2 D_n^{\text{C-flex}}}, \hspace*{\fill}
\end{multline}
where $d^{\text{C-fix}}$ is the level procured for the segment that is willing to pay up to the capacity market price cap, and $d_n^{\text{C-flex}}$ is the capacity procured for the $n$-th flexible segment. Our model includes segment-specific constraints similar to \eqref{eq:Consumer2}-\eqref{eq:Consumer3}.
\subsubsection{Market clearing conditions}

Finally, agent-specific decisions must comply with system-wide constraints. In particular, market-clearing conditions ensure a balance between supply and demand for both energy and CMs, and a constraint on the annual expected emissions ensures that the CO\textsubscript{2} emissions from electricity generation do not exceed the cap. The dual variables of these constraints, $\lambda^E_{\omega, t}$, $\lambda^C$, and $\lambda^{\text{CT}}$, can be interpreted as market prices (weighted with scenario probabilities and time interval durations). 
\begin{subequations}
\label{eq:SystemBalance}
\begin{gather}
\sum_{g }q_{\omega t, g} + \sum_{s } \left( q_{\omega t s}^{\text{dis}} - q_{\omega t s}^{\text{ch}} \right) \notag  \\
= d_{\omega t }^{\text{fix}} - d_{\omega t }^{\text{flex}} \quad \forall  t , \omega; \qquad \hspace*{\fill} (\lambda^{E}_{\omega t}) 
\label{eq:EnergyBalance} \\
\sum_{g }c^{\text{CM}}_g + \sum_{s}c^{\text{CM}}_s = d^{\text{CM}} ;  \qquad \hspace*{\fill} (\lambda^{\text{CM}}) 
\label{eq:CapacityBalance} \\
\sum_{g \omega t}  w_{\omega t} \delta_{\omega} q_{\omega t, g} \text{EF}_g \leq \Bar{E}^{\text{CT}} ; \qquad  \hspace*{\fill} (\lambda^{\text{CT}}) 
\label{eq:EmissionLimit}
\end{gather}
\end{subequations}

\subsection{Solution strategy}\label{subs:SolutionStrategy}

The market players' problems are convex quadratic with linear constraints, making their Karush–Kuhn–Tucker (KKT) conditions necessary and sufficient for optimality \cite{mokhtar_s_bazaraa_nonlinear_2005}. By combining KKT conditions and applying the integrability principle of variational inequalities \cite{steven_a_gabriel_complementarity_2013}, the solution to the central planner’s convex quadratic optimization problem equates to the equilibrium solution \cite{steven_a_gabriel_complementarity_2013}.
\subsection{Capacity accreditation}\label{subs:CC_estimation}

We estimate capacity credits using \textcolor{black}{a} marginal accreditation approach, supported by academic literature and increasingly adopted in industry \cite{zachary_integration_2022}. Credits are calculated for each generator $g\in\mathcal{G}$ and storage unit $s\in \mathcal{S}$, with storage evaluated across various durations $\zeta\in\mathcal{Z}$. Thus, the resource set for accreditation is $\mathcal{R} = \mathcal{G}\cup (\mathcal{S}\times\mathcal{Z})$.

 \textcolor{black}{The accreditation methodology builds upon the structure presented by Wang et al. \cite{wang_crediting_2022} to consider multiple storage durations. The methodology is similar to NESO's marginal EFC approach as it iteratively perturbs a \emph{baseline} portfolio  \cite{national_energy_system_operator_storage_2024}.\footnote{ \textcolor{black}{Our accreditation approach is closely aligned with the NESO EFC framework. However, rather than iteratively adjusting the firm capacity to find the quantity of perfectly firm capacity that restores the reliability metric to a target level, we directly evaluate the change in the specified reliability metric (EUE) resulting from a small perturbation of capacity. This approach is aligned with \cite{zachary_integration_2022} and avoids the need for iterative adjustments while preserving the marginal, risk-based interpretation of capacity credits, which is central to NESO’s method.}} In general, the \emph{baseline} portfolio can be obtained in two ways. One approach finds a \emph{baseline} portfolio by solving a welfare-maximizing problem that endogenizes the reliability value by considering the uncapped demand curve and the true WTP of consumers. Another approach finds a \emph{baseline} portfolio by embedding explicit reliability constraints (e.g., limits on EUE) directly in the planning problem with a price-capped demand curve (as in \cite{da_costa_reliability-constrained_2021}). In this article, we follow the former approach and obtain a \emph{baseline} portfolio ($\hat{c_r}$) that maximizes social welfare in the absence of price cap distortions, integrating the true consumers' WTP function (where prices can spike up to the VOLL). Dispatching the \emph{baseline} portfolio under the price-capped market, we determine the baseline $\text{EUE}_0$.} 
 
 Unlike in conventional models with perfectly inelastic demand, the definition of unserved energy is less straightforward when demand is price-elastic. In this work, we follow an approach similar to \cite{kaminski_impact_2021}. Considering all demand with WTP greater than or equal to the price cap as \textcolor{black}{fixed}, we record any shortfall of consumption with respect to that demand as unserved\footnote{Note that the price cap not only limits the energy price levels during scarcity but also removes part of the flexible demand, which would voluntarily adjust its level based on the energy price. Consistent with our assumption of efficient rationing, we treat this flexible segment as the first portion of demand to be shed once the price cap becomes binding.}. The unserved energy $l_{\omega t}$ is given by:
\begin{equation}\label{eq:ShedLoad}
    l_{\omega t} = \max(0,D_{\omega t}^{\text{fix}} + \frac{D^{\text{flex}}_{\omega t}}{\text{VOLL}} \cdot (\text{VOLL}-\text{PC})- d_{\omega t}^{\text{fix}}-d_{\omega t}^{\text{flex}})
\end{equation}

Then, the EUE is calculated as:
\begin{equation}\label{eq:ExpectedEnergyUnserved}
    \text{EUE} = \sum_{\omega t} w_{\omega t} \delta_{\omega} l_{\omega, t}
\end{equation}

\textcolor{black}{Subsequently, we perturb the \emph{baseline} portfolio ($\hat{c_r}$) by adding a hypothetical perfectly reliable reference generator with small capacity $\epsilon$ and re-dispatching the perturbed portfolio. We use $\text{EUE}_\text{ref}$ to note the expected unserved energy obtained from the redispatch of that perturbed portfolio. Similarly, for each resource $r$ (including storage at a fixed duration), we add $\epsilon$ of $r$ to the \emph{baseline} portfolio, redispatch the perturbed portfolio, and record $\text{EUE}_r$.} 

\textcolor{black}{Following the definition of capacity credit by Wang et al. \cite{wang_crediting_2022}, we calculate the marginal capacity credit as the EUE reduction caused by $\epsilon$ additional capacity of $r$, normalized by the EUE reduction caused by $\epsilon$ additional capacity of a perfectly reliable generator. The complete process is presented in Algorithm 1.}

\begin{algorithm}[!ht] \small \label{alg:capacity_credit}
\caption{Capacity Accreditation}
\begin{algorithmic}
\STATE \textbf{Input:} Optimal capacity mix $\hat{c}_r, \forall r \in R = G \cup (S \times E)$; system data; small capacity addition $\epsilon > 0$; duration set $\mathcal{Z} = \{\zeta_1, \zeta_2, \ldots\}$; model $\mathcal{M}$ composed of \eqref{eq:Gen},\eqref{eq:Consumer},\eqref{eq:SystemBalance}, with a price-capped energy demand function.
\STATE \textbf{Output:} Capacity credit $\text{CC}_r, \forall r \in R$

\STATE \textbf{Step 1: Compute baseline $\text{EUE}$}
\STATE Fix $c_r \leftarrow \hat{c}_r, \forall r \in \mathcal{R}$
\STATE \textcolor{black}{ Fix $e_{s}^{\text{init}} \leftarrow \hat{e}_{s}^{\text{init}}, \forall s \in \mathcal{S}$}
\STATE Solve model $\mathcal{M}$ to obtain baseline $\text{EUE}_0$

\STATE \textbf{Step 2: Perturb capacity for perfect generator}
\STATE Add $\epsilon$ of a perfectly flexible, reliable \text{ref}erence generator: $c_{\text{ref}} \leftarrow \epsilon$
\STATE Solve model $\mathcal{M}$ to obtain $\text{EUE}_{\text{ref}}$

\STATE \textbf{Step 3: Perturb capacity for each resource}
\FOR {$r \in R$}
    \STATE \textbf{if} {$r \in S \times \mathcal{Z}$} \textbf{then} Fix duration of storage $r$ to $\zeta \in \mathcal{Z}$
    \STATE Add $\epsilon$ of resource $r$: $c_r \leftarrow \epsilon$
    \STATE Solve model $\mathcal{M}$ to obtain $\text{EUE}_r$
    \STATE Compute capacity credit: $\text{CC}_r = \frac{\text{EUE}_0 - \text{EUE}_r}{\text{EUE}_0 - \text{EUE}_{\text{ref}}}$
\ENDFOR
\end{algorithmic}
\end{algorithm}

\textcolor{black}{Note that Algorithm 1 estimates capacity credits that are valid for the baseline portfolio. To capture the portfolio-dependent nature of marginal capacity credits, best practice recommends ELCC and EFC (hyper) surfaces \cite{mantegna_electric_2025, ming_resource_2025}. Depending on the application, capacity credit surfaces are constructed by estimating credits for various portfolios, or the surfaces are explored ad hoc by considering the portfolio obtained with the latest credits. Given the article's analytical focus, the latter approach is suitable. In our case study, as we show later in Section IV.C, the obtained and baseline portfolios are relatively similar, negating the need for iteratively adjusting the capacity credits.}

\subsection{Capacity market parameters}\label{subs:CapacityMarketParamerters}

\textcolor{black}{The capacity market WTP function is constructed as a piece-wise linear curve passing through  anchor points specified in the CM rules. The rules determine the coordinates of the anchor points as a function of the system-wide net CONE and capacity target $\text{CT}$. Following \cite{bothwell_crediting_2017}, we calculate the capacity target $\text{CT}$ as the total accredited capacity of the \emph{baseline} portfolio. The installed capacities of that portfolio are obtained from the welfare-maximising \textit{EOM VOLL} run, and denoted $\hat{c}_g$ for generators and $(\hat{c}_s^{P},\hat{\zeta}_s)$ for storage (power and duration):
\begin{equation}\label{CT_definition}
    \text{CT} =  \sum_{g} \text{CC}_g\, \hat{c}_g \;+\; \sum_{s} f^{\text{CC}}(\hat{\zeta}_{s})\, \hat{c}_{s}^{P}.
\end{equation}
This ensures that the capacity target is \textcolor{black}{consistent with the \emph{baseline} portfolio} used for marginal accreditation, with storage credits evaluated at the equilibrium duration $\hat{\zeta}_s$.}

Net CONE is defined as the difference between a technology’s annualised fixed and operational costs and its expected net revenues from the energy market, including ancillary services where applicable.  \textcolor{black}{For any resource $r$, let $\text{NC}^{\text{uc}}_r$ denote the net CONE per installed MW calculated when the \emph{baseline} portfolio is dispatched under the price cap. For generators and storage, this is given by:
\begin{subequations}
\label{eq:NetCONE}
\begin{gather}
\text{NC}^{\text{uc}}_g
=
\Bigg[
\hat{c}_g\left(\widetilde{I}_g + I_g^{F}\right)
-
\sum_{\omega,t}
w_{\omega t}\,\delta_\omega
\hat{q}_{\omega t g}
\left(
\hat{\lambda}^E_{\omega t}\right. \notag \\
\left.-C_g^{V}-\hat{\lambda}^{\text{CT}} \text{EF}_g
\right)
\Bigg]
\frac{1}{\hat{c}_g},
\quad \forall g \in \mathcal{G}, \label{eq:NetCONE_G} \\
\text{NC}^{\text{uc}}_s
=
\Bigg[
\hat{c}_s^{P}\left(\widetilde{I}_s^{P} + I_s^{P.F}\right)
+
\hat{c}_s^{E}\left(\widetilde{I}_s^{E} + I_s^{E.F}\right)
\notag \\
-
\sum_{\omega,t}
w_{\omega t}\,\delta_\omega
\left(
\hat{q}^{\text{dis}}_{\omega t s}
-
\hat{q}^{\text{ch}}_{\omega t s}
\right)
\hat{\lambda}^E_{\omega t}
\Bigg]
\frac{1}{\hat{c}_s^{P}},
\quad \forall s \in \mathcal{S}. \label{eq:NetCONE_S}
\end{gather}
\end{subequations}
}

Traditionally, CMs estimate net CONE for a reference technology $\text{NC}_{\text{ref}}$, typically a flexible conventional generator whose accredited capacity is based on its availability. We follow industry practice and use CCGT as the reference generator, assuming, for demonstration purposes, its perfect availability and flexibility. 

{\color{black}From an economic perspective, a well-designed CM with a single clearing price should ensure long-run cost recovery. This means that the capacity price per accredited MW should compensate for the missing revenues created by the price cap \cite{zachary_integration_2022,zuo_revisiting_2025}. We assume that a reference generator with a known capacity credit is actively present in the mix. In our setting, this reference generator is CCGT-CCS, which is taken to be perfectly available and flexible so that its accredited capacity equals its installed capacity (i.e., its capacity credit is 1). The long-run break-even condition for this reference generator implies that the capacity market price $\hat{\lambda}^{\text{CM}}$ equals its net CONE per installed MW, $\hat{\lambda}^{\text{CM}} = \text{NC}_{\text{ref}}$. The long-run break even condition for technology $r$ provides us with the  \emph{missing money-based} capacity credit $\text{CC}_r^{\text{MM}}$:
\begin{gather}
    \hat{\lambda}^{\text{CM}}
    = \frac{\text{NC}^{\text{uc}}_r}{\text{CC}^{\text{MM}}_r}
    \;\;\xRightarrow{\hat{\lambda}^{\text{CM}} = \text{NC}_{\text{ref}}}\;\; \notag \\
    \text{CC}_r^{\text{MM}}
    = \frac{\text{NC}^{\text{uc}}_r}{\text{NC}_{\text{ref}}}
    \quad \{\forall r\in \mathcal{R} \mid \hat{c}_r>0\}, 
\end{gather}

The \emph{missing money-based} capacity credit is the level that would allow each technology $r$ in the \emph{baseline} portfolio to recover the money that it is missing due to price cap distortions on short-term dispatch and price signals. Therefore, comparing marginal EFCs to $\text{CC}^{\text{MM}}$  indicates whether the capacity credits are at the level required to restore long-run cost recovery of the \emph{baseline} capacity mix. It is worth noting, though, that the \emph{baseline} capacity mix does not necessarily yield the maximum social welfare because price caps distort incentives for its optimal dispatch. 

Ref. \cite{zachary_integration_2022} proves that marginal EFCs are optimal under multiple assumptions about the power system structure and its operations. In particular, \cite{zachary_integration_2022} assumes that (1) storage and other flexible resources are dispatched to minimise expected unserved energy; (2) storage can be fully recharged between scarcity events, which are assumed to be well separated chronologically; and (3) risk metrics behave smoothly under marginal variations in capacity. In model $M$, where demand is elastic up to the price cap and storage dispatch is optimised for energy price arbitrage, these conditions may not hold.\footnote{\textcolor{black}{Solution to model $M$, while maximising the market surplus from dispatch, may not be EUE minimising. In our framework, this assumption may not hold as the model will prefer to shed load rather than charge storage at prices higher than $\text{PC}\cdot\eta_s^{\text{ch}}\cdot\eta_s^{\text{dis}}$. With price-elastic demand, the system can reduce flexible load up to this threshold, creating spare capacity to charge storage. An EUE-minimising schedule would therefore continue to charge storage or curtail flexible demand beyond the level implied by consumers’ willingness to pay, whereas the market surplus-maximising model will not. This behaviour appears in cases with low emission caps, where the effective marginal cost of gas generation becomes high.}}  Additionally, unlike \cite{zachary_integration_2022}, which assumes fixed demand, our model incorporates price-elastic demand, a setting that has not yet been examined in the context of EFC optimality with storage.}

\subsection{Experiment design} \label{subs:ExperimentDesign}
For our simulations, we use the model presented in \ref{subs:Model} in two distinct market structures: an energy-only market (\textit{EOM}), where CM-related variables and constraints are omitted, and a combined energy and CM (\textit{E+CM}). Each structure can be run with different energy demand functions, allowing the price to spike to VOLL or be capped at the price cap, set at an administratively defined level significantly lower than VOLL. When prices are allowed to spike to VOLL, the amount of the price-elastic demand is $D^{\text{flex}}_{\omega,t}$. With the price cap, the width of the linear part of the demand curve reduces by $D^{\text{flex}}_{\omega,t}\cdot\frac{\text{VOLL} - \text{PC}}{\text{VOLL}}$  and the flat part of the demand curve increases in width by the same amount.

\textcolor{black}{We simulate three main cases}: \textit{EOM VOLL} is an energy-only market \textcolor{black}{with prices} rising to VOLL; \textit{EOM PC} is an energy-only market with WTP capped at price cap; \textit{E+CM} is an energy+capacity market with WTP \textcolor{black}{for energy} capped at price cap. \textcolor{black}{When the price cap is applied, the welfare-maximizing mix is no longer financially feasible due to limited price spikes during scarcity, which curb the cost recovery. To identify financially viable capacity mixes in the presence of a cap on energy prices, we solve the \textit{E+CM} and \textit{EOM PC} models, which include and lack a capacity market, respectively.}

For comparative and accreditation purposes, we also run a case denoted \textit{EOM PC opt. mix}, where the price is capped, but the capacity mix is fixed to the welfare-maximising mix from \textit{EOM VOLL} (\textcolor{black}{also called \textit{baseline} portfolio}). The \textit{EOM PC opt. mix} model provides us with the baseline level for $\text{EUE}_o$, which is used to calculate capacity credits. 

\subsection{Performance metrics} \label{subs:PerformanceMetrics}

We use two primary performance metrics to evaluate the effectiveness of the \textit{E+CM} model. The first metric is the social welfare (SW), which is calculated as the difference between the consumer benefit \eqref{eq:Consumer1} and capital and operating costs.  By construction, the SW  of the \textit{EOM VOLL} run will be the highest because in that case, the objective function of the optimisation problem is identical to the SW metric. In the other runs, the SW loss measures the \textcolor{black}{difference in SW between that run and the \textit{EOM VOLL} run}. The second metric is EUE, as defined earlier in \eqref{eq:ExpectedEnergyUnserved}. \textcolor{black}{We also discuss how different the \textit{E+CM} and welfare-maximising capacity mixes are}. 

In this methodological framework, three factors can contribute to social welfare inefficiencies in the \textit{E+CM} market. First, \textcolor{black}{the shape of the demand curve for the capacity market is administratively set} \cite{zhao_constructing_2018}. \textcolor{black}{Even if the shape is correct,} calibrating CM parameters is challenging, and estimation errors in capacity credits or net-CONEs can impact the results. Second, imposing price caps narrows price spreads, disrupting storage operational patterns. Third, the unserved demand may be inefficiently allocated between periods as the model assigns a constant value of price cap to every MWh unserved. We evaluate SW according to the true WTP function, following Joskow and Tirole \cite{joskow_reliability_2007} in assuming efficient (non-price) rationing so that customers with the highest WTP are prioritised, but this means the average per-MWh cost of unserved energy rises with the amount unserved in that period. Under the alternative assumption of random rationing, the welfare cost of every MWh unserved would be equal, but the total cost of any given reduction in load would be higher than if it was targeted at the lower-WTP customers.

The latter two effects above relate solely to the introduction of a price cap, and can be assessed by running the \textit{EOM PC} case with optimal mix obtained from \textit{EOM VOLL} (\textit{EOM PC opt. mix.} mentioned earlier). To isolate these effects, we need to obtain the optimal true WTP-aware discharge schedule for the storage and remove the effects of true WTP inobservability on the EUE distribution, i.e., isolating the third effect. First, we fix the generation and charging decisions resulting from the original run. Subsequently, we replace the price-capped WTP with the true WTP, allowing it to reach the VOLL. Finally, we redispatch the mix, with only the discharging variables unfixed, which will now be scheduled to maximise true SW. Because the charging decisions of the storage are fixed, the schedule will only change discharging to better allocate the EUE across the time steps. Through this adjustment, we are left with only inefficiencies related to the first two effects, which are relevant to investigation in this paper. The third is out of scope as it relates to how to incentivize effective integration of demand-side flexibility under a price cap. Hence, in this paper, we assume that another scheme has successfully integrated demand-side flexibility to effectively distribute unserved energy.

\section{Case study setup}\label{subs:CaseStudySetup}

The model is applied to a greenfield expansion (i.e., assuming no existing capacity). 40 years (1980-2019) of weather-driven demand profiles and VRE availability are based on the Great Britain system. The average PV and wind (onshore) capacity factors are, respectively, 11\% and 29\%. The peak demand (i.e., $\text{max}_{\omega t}D_{\omega t}^{\text{fix}}$) is set to 100 MW and occurs in winter. The flexible demand is assumed to be 2MW for all time steps.  We simulate 3,600 time intervals of varying duration per year, identified through the chronological clustering technique of \cite{pineda_chronological_2018}, which is particularly suitable for modelling LDES. This results in a total of 144,000 (3600$\cdot$40) time steps. Based on the GB market, VOLL and price cap are set to  \$20,300/MWh and \$7,549/MWh\footnote{\color{black}{For VOLL it derives from £49/kW/year net CONE estimate with reliability standard of 3 hours LOLE \cite{noauthor_methodology_2025}, yielding VOLL of 16,333£/MWh. Energy price cap is assumed £6000/MWh \cite{noauthor_trading_2024}.}}, respectively.  Five technologies are considered: solar, wind, battery, LDES, and CCGT with CCS. Their techno-economic parameters are presented in Table \ref{tab:table1}. While in this paper, we keep a technology-agnostic perspective on LDES, we use parameters that are most reflective of the future hydrogen storage, obtained from review studies \cite{hunter_techno-economic_2021, albertus_long-duration_2020}. For the remaining technologies, we use NREL's ATB Database \cite{nrel_annual_2025}.

We study a diverse set of near-zero emission systems, which we identify by varying annual EI limits from 15 to 0.1 gCO\textsubscript{2}/kWh\footnote{\color{black}{All EI scenarios except 0.1\,gCO\textsubscript{2}/kWh align with recent \emph{Future Energy Scenarios} \cite{eso_future_2024}. The 0.1\,gCO\textsubscript{2}/kWh case is included primarily to illustrate systems with an even higher reliance on long-duration energy storage.}}. The model includes emission caps, which are calculated as the product of EI and the expected annual demand  ($\sum_{t \omega}[w_{\omega t}\delta_\omega (D_{\omega  t}^{\text{fix}}+ D_{\omega t}^{\text{flex}})]$). Assuming 90\% CCS efficiency, CCGT with CCS emits 37.8 gCO\textsubscript{2}/kWh. LDES is excluded from the optimal mix without an emission cap, concurring with previous research on the need for technological advancements \cite{sepulveda_design_2021}.

We construct the demand curves for the CM as follows. Following GB CM guidance \cite{national_energy_system_operator_auction_2024}, the curve has two downward-slopping segments, and their breakpoints are: (0.965\textbullet$\text{CT}$, 1.5\textbullet $\text{NC}_{\text{ref}}$), ($\text{CT}$, $\text{NC}_{\text{ref}}$), (1.035\textbullet$\text{CT}$, 0). The target capacity $\text{CT}$ is established as defined in \ref{subs:CC_estimation}.

\begin{table}[!htbp]
\centering
\caption{Techno-economic Parameters of Considered Technologies}
\label{tab:table1}
\setlength{\tabcolsep}{2pt}
\scriptsize
\begin{tabular}{lccccc}
\hline
\textbf{Parameter} & CCGT CCS & Solar & Wind & Battery & LDES \\ \hline
Power Capex (\$/kW) & 2500 & 895 & 1335 & 306 & 2000 \\
Energy Capex (\$/kWh) & - & - & - & 223 & 10 \\
FOM (\$/kW-yr) & 27 & 15 & 28 & 7.6 (5.6$^*$) & 40 (0.1$^*$) \\
Fuel cost + VOM (\$/MWh) & 40 & 0.5 & 0.5 & 0.5 & 0.5 \\
Charge Eff. (\%) & - & - & - & 92 & 60 \\
Discharge Eff. (\%) & - & - & - & 92 & 50 \\
Lifetime (years) & 30 & 30 & 30 & 15 & 30 \\
WACC (\%) & 7.1 & 6.2 & 6.2 & 7.1 & 7.1 \\ \hline
\end{tabular}

\vspace{0.3em}
\noindent\scriptsize{$^*$ FOM costs per MWh of storage energy capacity; WACC is Weighted Average Cost of Capital (WACC) used to annualize capital costs.}
\end{table}

\section{Results}

This section presents the results of the CM analysis, evaluating the \textit{E+CM} design against the \textit{EOM} benchmark across varying EI scenarios. Organised into five subsections, it examines the missing money problem, capacity credit estimation, the CM effectiveness and its effects on LDES, and sensitivity to accreditation inaccuracies.

\subsection{The missing money assessment}

\begin{figure}[!htbp]
\centering
\includegraphics[width=\columnwidth]{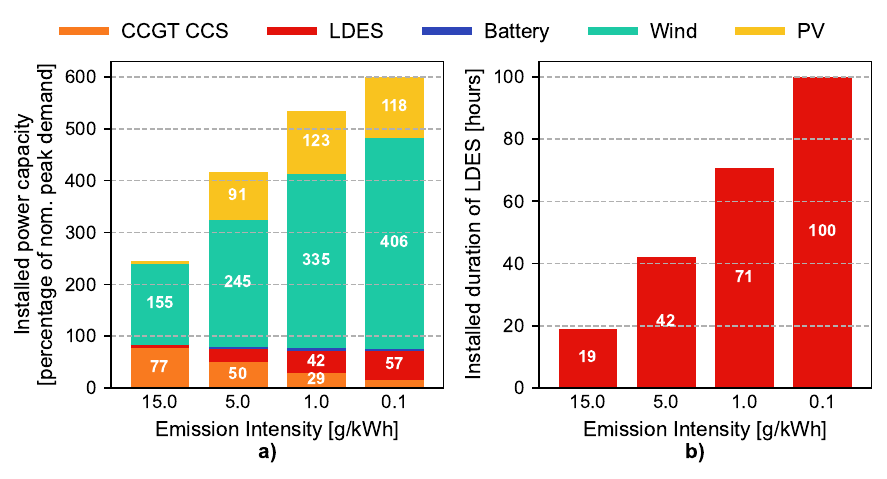}
\caption{Capacity results for the \emph{EOM VOLL} simulations. Panel a shows the installed power capacity as a percentage of peak demand, while panel b shows the installed duration of LDES. In the 15 gCO\textsubscript{2}/kWh case, no battery capacity is part of the welfare maximising mix, which is why the battery bar does not appear in this cluster and does not appear in subsequent results figures.}\label{fig:installed_capacity_voll}
\end{figure}

We derive a benchmark mix from \textit{EOM VOLL} simulations for each EI case. Fig.~\ref{fig:installed_capacity_voll} shows the capacity mix and LDES duration. Tighter EI targets reduce CCGT-CCS capacity while increasing wind, solar, and LDES. Batteries play a minor, auxiliary role\footnote{We do not model ancillary services, where batteries are expected to be essential. \color{black}{Furthermore, the chronological clustering we use may underestimate installed capacity of short-duration storage systems in large-scale case studies \cite{mannhardt_accurately_2025}.}}. Despite technological changes, non-VRE capacity remains near 80\% of peak demand, as firm capacity needs slowly decline with VRE expansion. Fig.~\ref{fig:installed_capacity_voll}b concurs with \cite{cebulla_how_2018}, with LDES duration rising sharply as decarbonisation deepens.

\begin{figure}[!htbp]
\centering
\includegraphics[width=\columnwidth]{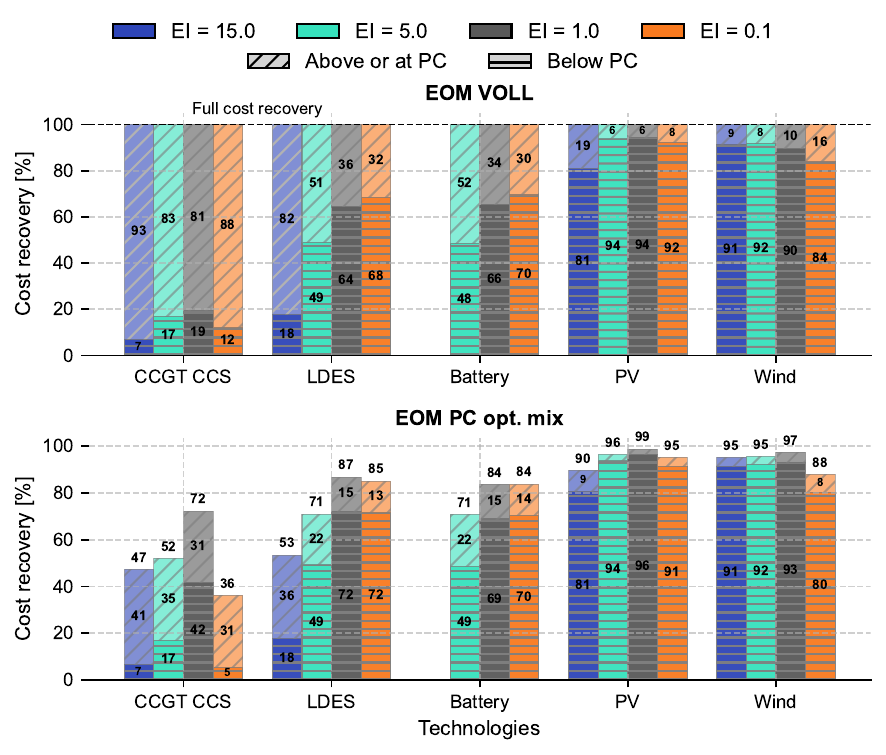}
\caption{Cost recovery per technology for \textit{EOM VOLL} and \textit{EOM PC opt. mix.}}
\label{fig:missing_money_plot}
\end{figure}

Fixing capacities at the levels from Fig. \ref{fig:installed_capacity_voll}, we dispatch resources under a price cap to estimate how much money each technology is missing to break even. Fig. \ref{fig:missing_money_plot} shows net revenues from the energy market per technology for two sets of time steps that have energy prices \textit{Below} and \textit{Above or at} price cap, respectively. For storage, we estimate the cost of energy discharged based on the average charging cost of stored energy at the time of discharge. Appendix 1 describes its calculation.

CCGT-CCS recovers more than 80\% of its costs during intervals with prices above the price cap in all EI cases. In contrast, storage recovers between 30–80\% in these intervals, with this fraction decreasing as emission limits tighten. This trend translates into a diminishing missing money problem for storage. Lower charging prices, driven by higher VRE generation and curtailment, result in reduced storage costs. At the same time, rising carbon costs increase the SRMC of CCGT-CCS, raising discharging prices when CCGT-CCS is the marginal producer, and thereby widening arbitrage spreads. We also observe that, unlike systems without storage, the introduction of the price cap not only reduces the \textit{Above or at} price cap net revenues, but also affects net revenues during hours when prices are below the price cap, highlighting the intertemporal nature of the missing money problem recently emphasized by \cite{visvesvaran_market-based_2025}. VRE technologies recover a large share of their costs in all cases and are less exposed to missing money. In contrast, CCGT-CCS remains the most vulnerable, as caps suppress scarcity rents that would otherwise support cost recovery.

\vspace{-1\baselineskip}

{\color{black}
\subsection{Capacity credit estimates} \label{sec:CapacityCreditEstimationResults}

We estimate the marginal CCs presented in Fig.~\ref{FIG: CapacityCredits} following Algorithm~1. For storage, we compute up to 56 marginal CCs, one per ES duration. We fit a four-segment piecewise linear curve to these estimates; the fitted curves closely follow the point estimates, with an $R^2$ exceeding 0.99 in all EI cases, confirming the appropriateness of this approach. The estimated CC curves are monotonically increasing and concave. 

Fig.~\ref{FIG: CapacityCredits} illustrates a critical dynamic: marginal CC estimates are highly sensitive to the underlying resource mix. In general, storage requires longer durations to achieve full accreditation as EI limits tighten, and for certain portfolios, full accreditation may not be achievable at all. This is because storage needs to shift larger amounts of energy as dispatchable capacity reduces. As the amount of `energy to shift' increases, the opportunities to charge (during hours with abundant VRE or other economical supply) and discharge (during hours of scarcity) become more chronologically separated, and the series of `scarcity events' (with price at PC) includes events more chronologically close to each other. When substantial time has lapsed from the previous scarcity event, system storage would be fully charged at the beginning of a `scarcity event'. However, when `scarcity events' start shortly after another event ended, system storage would likely go from empty to empty without an opportunity to fully charge in between. In that case, the system would appear to be \emph{charging-constrained}\footnote{\color{black}{We refer to storage as \emph{charging-constrained} when its adequacy contribution is limited by insufficient opportunities to recharge between chronologically-close scarcity events.}}. Under \emph{charging-constrained} conditions, the technologies that can most effectively shift energy have a higher capacity credit. In our test system, this is the case for batteries, which have a higher round-trip efficiency than LDES. This difference in capacity credits between storage technologies with the same duration is a novel observation in our analysis. It reflects that, as the system relies more on storage for adequacy,  round-trip efficiency becomes a more material driver of the adequacy contribution of storage, as it is directly linked with its ability to effectively use opportunities to charge between scarcity events.

\begin{figure}[!htbp]
\centering
\includegraphics[width=\columnwidth]{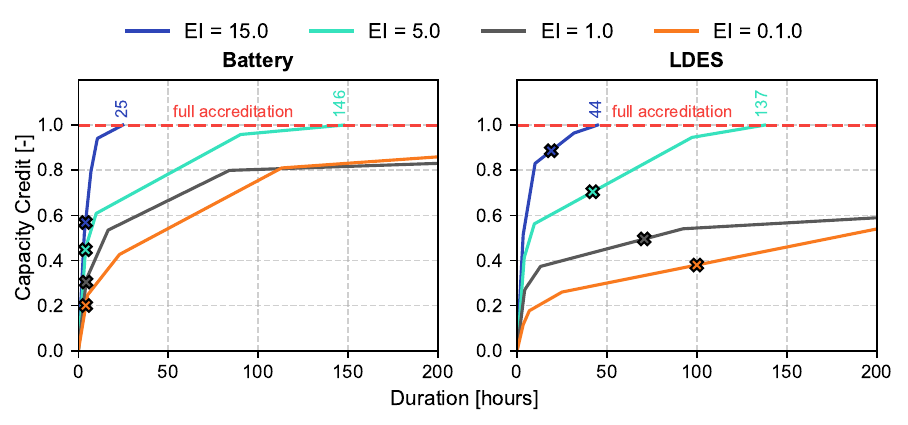}
\caption{Marginal capacity credit estimates as a function of storage duration. Markers indicate credit for duration obtained from \textit{EOM VOLL} simulations. }
\label{FIG: CapacityCredits}
\end{figure}

}

{\color{black}
The \emph{charging-constrained} conditions complicate the marginal accreditation calculations in two ways. {\color{black}First, when a marginal storage unit is added to the baseline portfolio, it will reduce EUE during \emph{charging-constrained} hours if the constraint was due to insufficient charging power capacity, but not if there was insufficient surplus energy available before the next shortage period.} Second, when a marginal generator is added to the baseline portfolio, it can reduce EUE not only directly in shortage hours, but also indirectly by creating additional surplus energy between scarcity events that storage can shift into a shortage period.

Fig.~\ref{FIG: CCGT_CCG_EUE_change} illustrates the second mechanism by comparing the marginal EUE reduction from adding a small increment of CCGT-CCS in Step~2 of Algorithm~1 under two dispatch setups: (1) \textit{unconstrained} storage dispatch; (2) \textit{existing ES charging fixed}. The latter setup fixes storage charging at the level obtained from the price-capped dispatch of the baseline portfolio. When the EUE reduction is larger under (1), the small increment of CCGT-CCS might be used to charge storage. The difference between the EUE reductions under the two setups is most pronounced in the 1~gCO\textsubscript{2}/kWh case, indicating more severe charging-constrained conditions. The differences are small in other EI cases. For high EIs, the charging-constrained conditions are less severe. For the 0.1~gCO\textsubscript{2}/kWh case, the effective marginal cost of CCGT-CCS (carbon-inclusive) is economically unattractive for charging LDES, but still attractive for charging batteries. 

\begin{figure}[!htbp]
\centering
\includegraphics[width=\columnwidth]{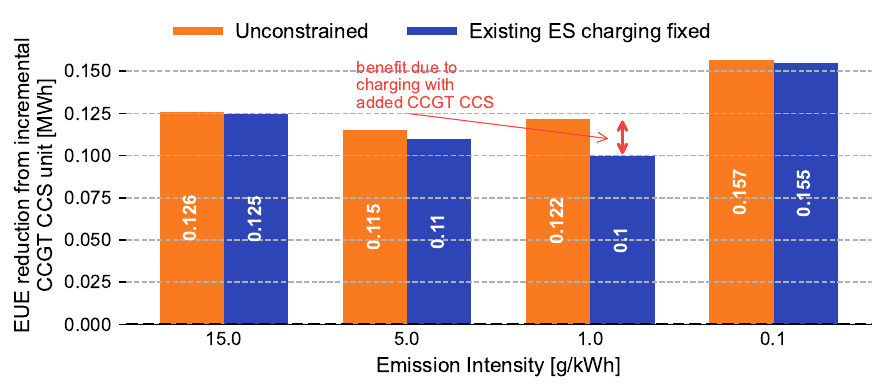}
\caption{CCGT-CCS marginal 0.01MW unit contribution to EUE reduction under alternative setups.}
\label{FIG: CCGT_CCG_EUE_change}
\end{figure}

As a result of these two mechanisms, under \emph{charging-constrained} conditions, the capacity credit estimated with Algorithm 1 is less linked with contributions of marginal technologies during scarcity hours and their missing money. For a generator, the estimated capacity credit also captures the generator's indirect contribution to reducing EUE, which is unlikely to be associated with missing money. For storage, the estimated capacity credit reflects a lower adequacy contribution due to stricter charging constraints for the marginal addition. Hence, the estimated capacity credits are expected to be lower than required for recovering missing money for storage or generators with smaller indirect contributions than the reference generator; and larger for generators with larger indirect contributions than the reference generator. This is indeed the result we observe in Fig. \ref{FIG: NetConesStorage}, where we compare the missing money-based credit to the EUE-based credit. 

{\color{black}In Fig. \ref{FIG: NetConesStorage}, the bars depict the corresponding net CONEs per (EUE-)accredited MW. Capacity markets have hitherto operated on the basis that, in equilibrium, the net CONE per accredited MW of installed resources should be identical. However, this is not the case for the baseline portfolio with EUE-based credits for our system. {\color{black}As well as} the dynamics described above related to storage being charging constrained, we identify two more mechanisms that could explain the differences.}

 \begin{figure}[!htbp]
\centering
\includegraphics[width=\columnwidth]{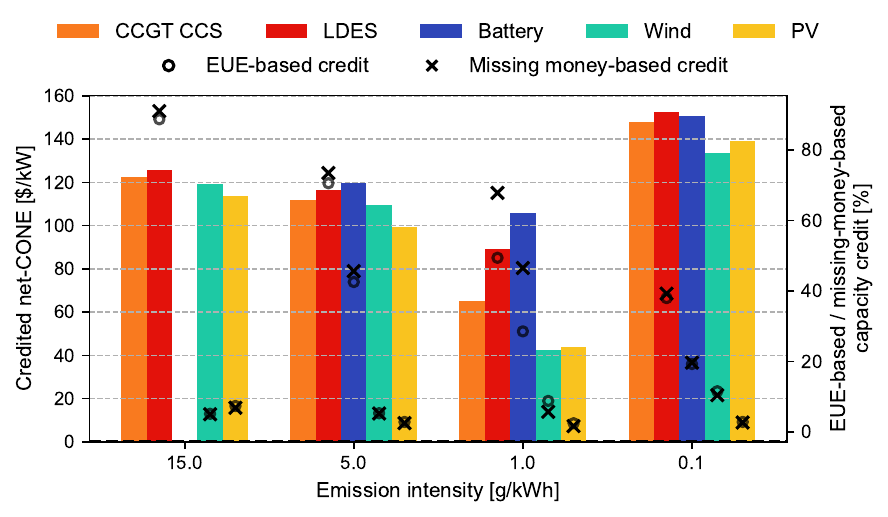}
\caption{Credited net-CONEs and capacity credits for technologies identified by \textit{EOM VOLL} simulations. Note for storage, the optimal duration is considered.}
\label{FIG: NetConesStorage}
\end{figure}

First, demand is assumed to be price-elastic above the price cap. With elastic demand, uncapped scarcity prices may vary between the marginal cost of the marginal generator and VOLL, so the missing money per MWh induced by a price cap is not constant but varies across hours. Therefore, if a resource's output is correlated with the level of missing money per MWh, the missing-money shortfall per MW of that resource can be relatively greater than its marginal contribution to reducing unserved energy. In our case, the output of storage is more correlated with the level of missing money per MW than the reference generator. Storage can be energy-constrained while the reference generator is not, and it discharges only when prices reach the shadow value of stored energy \cite{forsund_hydropower_2007, junge_energy_2022}. If this shadow value lies above the price cap, there will be intermediate scarcity hours in which reference generators incur relatively modest missing-money shortfalls, but storage does not, since it would not have discharged at those times. Prices can also exceed the shadow value of stored energy if storage is power-constrained while discharging, so that these maximal storage discharges are associated with the largest amounts of missing money per MWh. Indeed, in Table~\ref{tab:PriceStorageHours}, we observe that the higher the number of intermediate scarcity hours, the larger the differences in net CONE per accredited MW. 

Second, some of the missing money for storage is due to the lack of incentives for efficient dispatch in a price-capped environment. In a market without a price cap, storage may optimally charge at prices above $\text{PC}\cdot\eta_s^{\text{ch}}\cdot\eta_s^{\text{dis}}$ when the discharge value justifies doing so. However, with a price cap, such charging is no longer profitable because the cap truncates the maximum discharge price. Table~\ref{tab:PriceStorageHours} confirms that these `high-price' charging events occur in the uncapped dispatch, particularly for LDES under tighter EI targets. By changing these charging decisions and compressing spreads, the price cap can therefore reshape both storage schedules and energy prices outside scarcity hours, further supporting our discussion in Figure \ref{fig:missing_money_plot} that the missing-money distortion has an intertemporal character and is not confined to the scarcity interval. Due to these distorted incentives for dispatch, the baseline portfolio might not maximise the social welfare in an energy-plus-capacity market environment. 

{\color{black}These dynamics can be intertwined, and their relative importance depends on the underlying system, so they can shift whether (and how closely) EUE contribution aligns with the missing-money shortfall that a CM is intended to recover. Our results indicate that maintaining this alignment can be challenging in deeply decarbonised systems with high storage penetration and elastic demand, motivating more fundamental derivations of the sufficient conditions under which EFC-based accreditation restores long-run cost recovery, which we leave for future work. Following current industry practice, in the remainder of the results, we therefore proceed using the  EUE-based credits to evaluate how the capacity market performs under our case study assumptions.}

\begin{table}[!htbp]
\centering
\scriptsize
\caption{Average hours per year of specific price and storage-dispatch instances from EOM VOLL case.}
\label{tab:PriceStorageHours}
\setlength{\tabcolsep}{4pt}

\begin{tabular}{p{1.3cm} *{4}{>{\centering\arraybackslash}p{1.3cm}}}
\toprule
& \multicolumn{4}{c}{\textbf{Emission intensity (gCO$_2$/kWh)}} \\
\cmidrule(lr){2-5}
\textbf{Parameter} & \textbf{15.0} & \textbf{5.0} & \textbf{1.0} & \textbf{0.1} \\
\midrule

\multicolumn{5}{l}{\textbf{Intermediate scarcity price region:}  
$\text{PC} < \lambda^E_{\omega t} < \text{VOLL}$} \\
Hours   & 4.93 & 5.30 & 15.73 & 14.28 \\
\midrule

\multicolumn{5}{l}{\textbf{High price charging hours:}  
$q^{\text{ch}}_{\omega t s} > 0$ and $\lambda^E_{\omega t} > \text{PC}\cdot\eta_s^{\text{ch}}\cdot\eta_s^{\text{dis}}$} \\
Battery & - & 0.63 & 1.23 & 0.20 \\
LDES    & 0.38 & 1.10 & 7.50 & 8.93 \\

\bottomrule
\end{tabular}

\end{table}
}
 
\subsection{Effectiveness of energy plus capacity market}

Under a price cap, SW losses are caused by two factors.  First, unserved demand is inefficiently distributed among consumers due to a lack of price signals between the price cap and VOLL. Second, distorted price signals result in inefficient ES dispatch. In Table~\ref{TAB:PerformanceMetrics}, we show SW losses caused by the second factor for the \textit{EOM PC} and \textit{E+CM}. For the benchmark capacity mix dispatched under price cap, we present SW losses for two cases: when both factors are present (\textit{EOM PC opt. mix. ineff. dist.}) and when only the second factor affects results (\textit{EOM PC opt. mix.}). Under higher emission limits, welfare losses almost disappear when demand rationing is perfect, indicating that inefficient distribution of unserved energy is the primary reason for SW losses. However, for the  0.1gCO\textsubscript{2}/kWh, the welfare loss is high even under perfect demand rationing, indicating a substantial impact of the price cap on storage operations via reduced price spreads.

 \begin{table}[!ht]
\centering
\caption{Social Welfare and Unserved Energy for All Runs}
\label{TAB:PerformanceMetrics}
\setlength{\tabcolsep}{4pt}
\scriptsize

\textbf{Welfare Loss [\% TC in \textit{EOM VOLL}]}\\[3pt]
\begin{tabular}{c|cccc}
\toprule
\textbf{EI} 
& \begin{tabular}{@{}c@{}}EOM PC opt. mix.\\ineff. dist.\end{tabular}
& \begin{tabular}{@{}c@{}}EOM PC\\opt. mix\end{tabular} 
& E+CM 
& \begin{tabular}{@{}c@{}}EOM PC\end{tabular} \\
\midrule
15.0 & -0.0140 & -0.0003 & -0.0045 & -0.8536 \\
5.0  & -0.0521 & -0.0078 & -0.0156 & -0.7101 \\
1.0  & -0.0338 & -0.0114 & -0.0439 & -0.4253 \\
0.1  & -0.7876 & -0.6862 & -0.3776 & -1.6923 \\
\bottomrule
\end{tabular}

\vspace{1.2em}

\textbf{Unserved Energy [\% of total demand]}\\[3pt]
\begin{tabular}{c|cccc}
\toprule
\textbf{EI} 
& \begin{tabular}{@{}c@{}}EOM VOLL$^*$\end{tabular}
& \begin{tabular}{@{}c@{}}EOM PC opt. mix\end{tabular}
& E+CM 
& \begin{tabular}{@{}c@{}}EOM PC\end{tabular} \\
\midrule
15.0 & 0.0043 & 0.0043 & 0.0044 & 0.0143 \\
5.0  & 0.0034 & 0.0034 & 0.0036 & 0.0133 \\
1.0  & 0.0018 & 0.0019 & 0.0025 & 0.0088 \\
0.1  & 0.0027 & 0.0080 & 0.0080 & 0.0294 \\
\bottomrule
\end{tabular}

\vspace{0.4em}
\noindent\scriptsize{$^*$ For comparability, for the \textit{EOM VOLL} case we measure EUE relative to the fixed part of the true WTP and the 'cut-off' part of flexible demand above the PC.}

\end{table}

Under the \textit{E+CM} design, we observe welfare losses, but they are significantly lower than those observed with price cap but no CRM (\textit{EOM PC}).  For all cases with EI higher than 0.1gCO\textsubscript{2}/kWh, welfare losses are minimal (below 0.1\% of total cost) under the \textit{E+CM} design, yet growing with a tighter emission limit. However, in the 0.1gCO\textsubscript{2}/kWh case, the welfare loss is almost 0.4\% of total cost under the \textit{E+CM} design, which might seem high, but it is still roughly half the loss caused by the price cap alone (\textit{EOM PC opt. mix.}). In \textit{E+CM} design, players account for price cap distortions, and their optimal solutions differ from the benchmark mix, primarily in terms of reduced LDES capacity (-1.3 MW/-60 MWh). That mix is more efficiently dispatched, and the SW is closer to the optimal level under the \textit{E+CM} compared to a design that supports the benchmark capacity mix. At 1gCO\textsubscript{2}/kWh, losses are likely driven by the weaker relationship between missing money and EUE contributions. In terms of EUE, comparing \textit{EOM VOLL} and \textit{EOM PC opt. mix.}, we see that the impact of the price cap is substantial under the lowest EI scenario. EUE levels under \textit{E+CM} closely match the \textit{EOM PC opt. mix.} levels, except for the 1gCO\textsubscript{2}/kWh case, which again derives from weaker relationship between missing money and EUE contributions.

{\color{black}Table~\ref{tab:abs_changes} reports the change in installed capacities in the CM relative to the \emph{baseline} portfolio. The welfare losses observed in Table~\ref{TAB:PerformanceMetrics} stem from different investment and dispatch incentives in the E+CM vs. the EOM with VOLL case.  Under all cases for emission intensity, the E+CM portfolio has higher investment in CCGT-CCS and lower investment in LDES compared to the \emph{baseline} portfolio. This substitution effect is expected for competing technologies contributing to reliability. Similarly, we observe a substitution between wind and solar investments, with higher EIs having lower wind and higher solar, and lower EIs having higher wind and lower solar than the baseline portfolio. For batteries, we observe mixed results across EIs as they can complement renewables, particularly solar, but also act as a substitute for LDES. We also see that the \textit{E+CM} mix from 0.1gCO\textsubscript{2}/kWh case, which, in terms of welfare, performs better than the \textit{EOM VOLL} mix with price cap, results in lower LDES capacity and higher CCGT-CCS. It indicates that distortions caused by the price cap in the dispatch of storage generally favour conventional generation.  Overall, the differences between the \emph{baseline} portfolio and the \textit{E+CM} portfolio are relatively small. That's why we have not included capacity credit surfaces that account for the impact of the portfolio on capacity credits. This small difference is not guaranteed, and that's why systematic integration of EFC surfaces in a capacity market or resource adequacy process is recommended.
\begin{table}[!htbp]
\centering
\caption{Absolute Change in Installed Capacities Between E+CM and EOM VOLL Cases}
\label{tab:abs_changes}
\setlength{\tabcolsep}{2pt}
\scriptsize
\begin{tabular}{lccccc}
\hline
\textbf{EI} & \textbf{PV} & \textbf{Wind} & \textbf{CCGT CCS} & \textbf{Battery*} & \textbf{LDES*} \\ \hline
15.0 & 1.05 & -0.42 & 0.41 & -- / -- & -0.57 / -23.25 \\
5.0 & 1.14 & -0.39 & 0.14 & 0.45 / 2.39 & -0.70 / -46.48 \\
1.0 & -1.36 & 2.68 & 0.06 & -0.79 / -2.56 & -0.94 / 1.96 \\
0.1 & -2.79 & 0.83 & 0.20 & 0.05 / 4.14 & -1.32 / -60.66 \\ \hline
\end{tabular}

\vspace{0.3em}
\noindent\scriptsize{Absolute changes from EOM VOLL run. 
Positive values indicate higher capacity in the E+CM case. 

$^*$For storage technologies, values show the change in power capacity / energy capacity (MW / MWh).}
\end{table}

}

\subsection{CM impacts on prices and LDES net revenues}
Introducing a capacity market affects energy prices and LDES revenues. Table \ref{TAB:PriceCostMetrics} contains consumer cost and price volatility metrics. Tighter emission limits raise energy costs. Due to price caps and payments through CMs, energy prices in \textit{E+CM} are lower by 15–30 \$/MWh compared to \textit{EOM VOLL}, but capacity procurement costs, passed to consumers, offset these savings. Thus, consumer costs are similar in \textit{EOM VOLL} and \textit{E+CM}, indicating no extra financial burden for the CM.\footnote{Administrative and transaction costs, which could marginally impact overall costs, are not considered in this analysis.} CMs also lower energy price volatility, though this effect diminishes with stricter emission limits.

\begin{table}[!htbp]
\centering
\caption{Price and Cost Metrics for All Runs}
\label{TAB:PriceCostMetrics}
\setlength{\tabcolsep}{3pt}
\scriptsize
\begin{tabular}{c|ccc|cccc}
\hline
\multirow{2}{*}{\begin{tabular}[c]{@{}c@{}}Em. Intensity\\ {[}gCO$_2$/kWh{]}\end{tabular}} 
& \multicolumn{3}{c|}{\begin{tabular}[c]{@{}c@{}}EOM VOLL\end{tabular}} 
& \multicolumn{4}{c}{\begin{tabular}[c]{@{}c@{}}E+CM\end{tabular}} \\
\cline{2-8}
& $\mu_{\text{Price}}$ & $\sigma_{\text{CV}}$ & $\beta_{\text{EB}}$
& $\mu_{\text{Price}}$ & $\sigma_{\text{CV}}$ & $\kappa_{\text{CCC}}$ & $\beta_{\text{EB}}$ \\ \hline
15.0 
& 99 & 0.42 & 16 
& 80 & 0.24 & 19 & 16 \\
5.0  
& 122 & 0.52 & 18 
& 106 & 0.34 & 16 & 18 \\
1.0  
& 139 & 0.78 & 12 
& 129 & 0.66 & 10 & 12 \\
0.1  
& 156 & 1.26 & 8 
& 133 & 1.03 & 21 & 7 \\ 
\hline
\end{tabular}

\vspace{0.3em}
\noindent\scriptsize{
$\mu_{\text{Price}}$: Annual average energy price [\$/MWh].\quad
$\sigma_{\text{CV}}$: Coefficient of variation of annual energy price [–].\quad
$\kappa_{\text{CCC}}$: Average annual consumer capacity cost [\$/MWh].\quad
$\beta_{\text{EB}}$: Emission benefit from carbon credits [\$/MWh].\\
Note: The emission benefit ($\beta_{\text{EB}}$) derives from the shadow price of the emission constraint. It may not equate to a societal valuation of avoided emissions.
}
\end{table}

\begin{figure}[!htbp]
\centering
\includegraphics[width=\columnwidth]{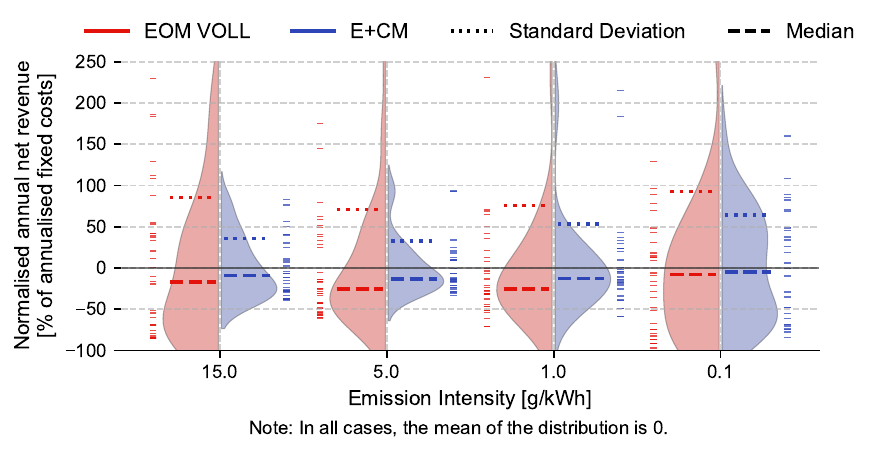}
\caption{Distribution of annual net revenues for LDES across EI cases.}
\label{FIG:NetRevenueDistribution}
\end{figure}

LDES breaks even on average but not annually, as value accrues during rare scarcity events. Fig. \ref{FIG:NetRevenueDistribution} shows the annual net revenue distributions for LDES. In all cases, the distribution is highly positively skewed, with more than half of the outcomes reflecting annual losses, but a few extreme cases of very high profits. For cases with tighter emission limits, the CM proves highly effective in reducing both net revenue volatility and the positive skewness, which is an essential consideration for risk-averse investors. However, this stabilising effect weakens as emission constraints tighten. 

\subsection{Sensitivity to capacity credit estimation inaccuracies}

\begin{figure}[!htbp]
\centering
\includegraphics[width=\columnwidth]{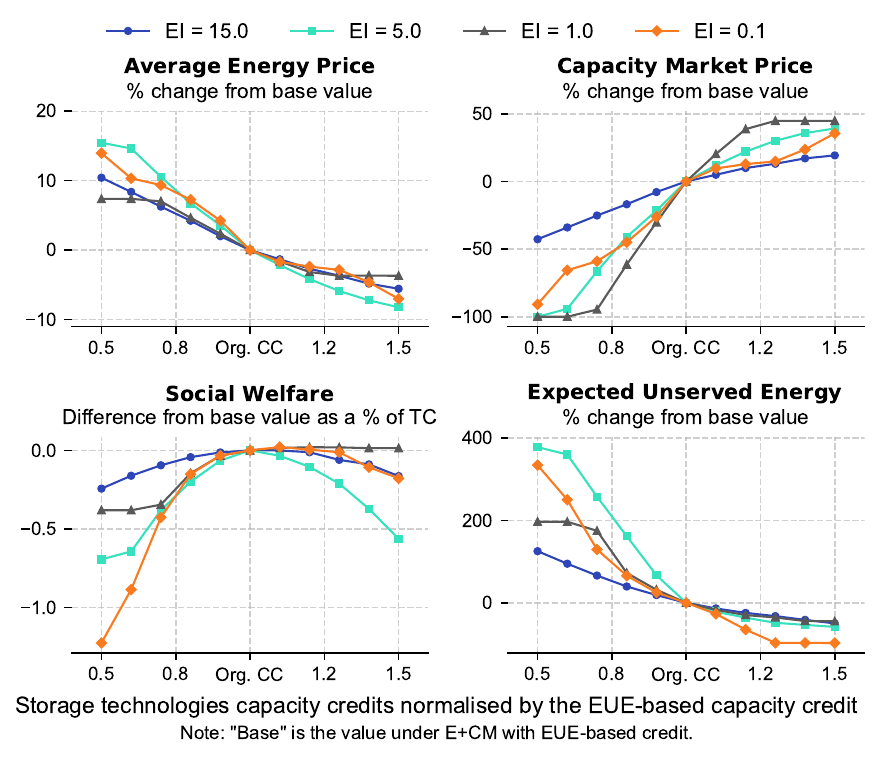}
\caption{Sensitivity of performance to storage credits.}
\label{FIG:scale_credits_sensitivity}
\end{figure}

We assess the impact of capacity credit estimation errors on market performance through a sensitivity analysis, adjusting the capacity credit of storage technologies to account for the dependence of CC on the resource mix and potential inaccuracies of the accreditation methodologies. For each simulation, we use the adjusted credits to estimate a new capacity target. We keep the net-CONE for the reference generator the same across all sensitivities. Fig. \ref{FIG:scale_credits_sensitivity} shows the change in the average energy price, capacity price, social welfare, and EUE. Under credit underestimation, the investment signal provided by the capacity market to storage weakens, leading to lower investment and tighter conditions in the energy market. As a result of lower capacity targets, capacity market prices reduce, and energy prices and EUE increase. The changes are opposite under credit overestimation. We also find an asymmetric effect on welfare, with underestimation causing larger losses. The SW marginal increases with overestimated credits for 1 and 0.1 gCO\textsubscript{2}/kWh cases. In the former case, this is due to weaker relationship between missing money and EUE contributions (per Section \ref{sec:CapacityCreditEstimationResults}), and in the latter, due to the mix that better negates the impact of the price cap on storage dispatch. 

\section{Discussion and conclusions}

This study investigates whether CMs can effectively address the problem of missing money, caused by price caps, and ensure system adequacy and high social welfare in deeply decarbonised electricity systems with high LDES penetration. Methodologically, we extend state-of-the-art equilibrium investment models by introducing continuous, duration-based accreditation for storage. By formulating capacity credit curves as piecewise linear functions of storage duration, we enable efficient integration of flexible-duration ES into an investment model. Formulations presented here are useful not only for market-based studies but also for reliability-constrained capacity expansion models with planning reserve margins. 

The analysis is performed for a case study with renewable and load profiles that resemble those of Great Britain. Simulations consider a range of CO\textsubscript{2} emission targets that yield varying levels of LDES integration. We find that price cap distortions result in a substantial missing money problem for LDES (approximately 15\% to 50\% of LDES costs are not recovered when a baseline mix identified by maximising SW is dispatched with energy price caps). The magnitude of the problem reduces under more decarbonized systems, as storage can capture higher spreads due to low charging costs during periods of abundant renewable supply and high discharging costs during periods with expensive carbon-emitting marginal suppliers.

We demonstrate that well-calibrated CMs can provide near-efficient investment signals, resulting in welfare outcomes comparable to those of an energy-only market without price caps. These results expand the work of Wang et al. \cite{wang_crediting_2022}, by showing that efficient outcomes are possible even when demand is flexible, storage duration is an investment decision, and ES makes up a significant share of installed capacity. In cases of deeper decarbonization, we find that the capacity market is less effective in restoring system security of supply, as measured by expected unserved energy, to the levels that would prevail without price caps. Two factors explain this result. \textcolor{black}{First, the relationship between missing money and marginal contributions to adequacy varies per technology, when demand is price-elastic, and storage is charging-constrained}. Second, in the presence of storage, energy price caps distort market outcomes beyond periods of scarcity. By capping discharging revenues, the price cap undermines incentives for charging in non-scarcity hours.  This suggests that conventional methods that only consider the difference between price cap and VOLL may be poorly suited to parameterize capacity markets in deeply decarbonized systems.

Our work provides a series of timely regulatory insights. While CMs can, in principle, remain an effective tool for ensuring adequacy, they are complex and sensitive to parameter values such as capacity credits. Nonetheless, CMs effectiveness can remain high in the medium term, provided that accreditation methods are transparent and predictable, giving investors greater certainty.  \textcolor{black}{In the longer term, capacity market parameterization methodologies may need to be refined to reflect how price caps, in systems with high storage penetration, affect revenue formation and reliability contribution not only during scarcity hours but also outside scarcity periods.} Given that the ability of CMs to reduce price and revenue volatility declines with deeper decarbonisation, complementary measures, such as risk-hedging government-backed contracts, are worth further investigation.

Future work on business models for LDES can extend the framework presented in this article by modeling risk-averse investors, considering relevant policy mechanisms and novel electricity market products that might be a good fit for LDES, and limiting the horizon within which LDES has foresight into future energy prices. \textcolor{black}{Performance penalties are another key design element of CRMs that merits deeper investigation, as their enforcement is complicated by the presence of \textit{charging-constrained} conditions. Furthermore, the framework can be extended to include network constraints and an explicit representation of the system operator. This could be implemented through zonal or nodal formulations that account for transmission limits and locational marginal pricing, with the system operator acting as an additional agent ensuring network feasibility of the energy schedule. Such extensions would allow the assessment of spatial adequacy and locational incentives under alternative market structures. Finally, an important avenue for future research is the development of more fundamental mathematical conditions under which marginal EFC-based accreditation restores long-run cost recovery and adequacy in systems with price-elastic demand and charging-constrained storage.}

\section*{Appendix 1}

To estimate storage net revenues per time step, we first calculate the time-varying book value of stored energy based on charging prices. For each scenario $\omega$ and time step $t$, the book value per MWh of stored energy, $\phi_{\omega,t,s}$, is updated as follows:
\begin{align}
\phi_{\omega t s} =\; & \left( \phi_{\omega, t-1, s} \cdot \left( e_{\omega, t-1, s}  - w_{\omega t} \cdot q^{\text{dis}}_{\omega t s} / \eta_s^{\text{dis}} \right) \right. \notag \\
& \left. +\; \lambda^E_{\omega t}  \cdot  w_{\omega t} \cdot q^{ch}_{\omega t s} \right) \cdot \left( 1 / e_{\omega t s} \right)  \quad \forall  \omega t s \hspace*{\fill}
\label{EQ: StorageValueBalance1}
\end{align}

We need to set the value of the initial stored energy. Since we constrain the initial and final energy levels to be equal, as if the model were continually cycling through its year of time periods, we impose a similar restriction on their values. We iterate the calculation of equation \eqref{EQ: StorageValueBalance1} until $\phi_{\omega,0,s} = \phi_{\omega,T,s}$, starting from  $\phi_{\omega,0,s} = 0.$

Considering this time-varying value as the cost of discharging, we estimate the net revenue from participation in storage in the energy market at each time step as follows:
\begin{equation}
\pi^{\text{op}}_{\omega t s} = q^{\text{dis}}_{\omega t s} \cdot \left( \lambda^E_{\omega t} - \phi_{\omega, t-1, s} / \eta_s^{\text{dis}} \right)  \quad \forall  \omega t s  \hspace*{\fill}
\end{equation}

\vspace{-0.5\baselineskip}
\printbibliography

\end{document}